\documentclass[pre,aps,showpacs,twocolumn,floatfix]{revtex4}
\usepackage[dvips]{graphicx}
\usepackage{amsmath}
\usepackage{amsfonts}
\usepackage{amssymb}

\begin{document}

\title{Influence of the  Barrier Shape on Resonant Activation}

\author{Bart{\l}omiej \surname{Dybiec}}
\email{bartek@zfs.if.uj.edu.pl}

\author{Ewa \surname{Gudowska-Nowak}} 
\email{gudowska@th.if.uj.edu.pl}

\affiliation{Marian~Smoluchowski Institute of Physics,\\
 Jagellonian University, Reymonta~4, 30--059~Krak\'ow, Poland}

\date{\today}

%
%
\begin{abstract}
The escape of  a Brownian particle over a dichotomously fluctuating barrier is investigated
for various shapes of the barrier. The problem of resonant activation is revisited with the 
attention on the effect of the barrier shape on optimal value of the mean escape time in the
system. The characteristic features of resonant behavior are analyzed for 
situations when the barrier switches
either between different heights representing erection of a barrier and
formation of a well, respectively, or it proceeds through ``on'' and 
 ``off'' positions.
\end{abstract}


\maketitle

%
%
\section{Introduction}

Nonequilibrium systems driven by noises are known
to display plentitude of noise-induced phenomena \cite{lef,kam}.
Among those, stochastic resonance is manifested when the response of a nonlinear system to a
signal is enhanced by the presence of noise. Another resonant phenomenon can be observed for
thermally activated surmounting of a potential barrier \cite{doe} with a randomly fluctuating shape.
For certain values of characteristic parameters of the noise, the transport over the barrier is facilitated, \textit{i.e.}, 
the mean escape time of a particle exhibits a minimum  as a function of the parameters of the
barrier fluctuations. Such a problem of thermally activated escape within randomly fluctuating
potentials occurs in a wide variety of contexts with examples ranging from molecular dissociation
in strongly coupled chemical systems \cite{mad} or the model dynamics of the dye laser \cite{fox} 
to selective ion pumps
in biological membranes \cite{lee}. In every case activation happens due to
thermal fluctuations and after classical Kramers theory \cite{kra} can be expressed by
an Arrhenius dependence for the mean lifetime of the metastable
state, $W\sim\exp(\Delta G/k_BT)$, where $\Delta G$ stands for the activation 
energy. The strong effect of the surroundings (``environmental noises'') can be
readily understood since even small variations $\delta G$  in $\Delta G$ will
greatly affect $W$ provided $\delta G >k_B T$. For a complex nonequilibrium 
system the potential experienced by the Brownian particle cannot, in general,
be regarded as static but rather, as influenced by random fluctuations whose
characteristic time scale may be comparable with the duration of the
diffusion over the barrier. Random effects of the
environment
can thus be envisioned as  barrier alternating processes \cite{goel} that can modulate 
the escape kinetics in the system.
If the barrier fluctuates extremely slowly, the mean first passage time (MFPT) to
the top of the barrier is dominated  by those realizations for which the
barrier starts in a higher position and thus becomes very long. The barrier
is then essentially quasistatic throughout the process. At the other extreme,
in the case of rapidly fluctuating barrier, the mean first passage time is
determined by the ``average barrier''.
 For some particular correlation time of the barrier
fluctuations, it can happen however, that the mean kinetic rate of the process 
 exhibits an extremum
\cite{doe,iwa,bork} that is a signature of resonant tuning of the system in response
to the external noise.

In this communication we present numerical results for the mean first passage time over a
fluctuating barrier for the model system with an \textit{arctan} potential barrier
subject to dichotomous fluctuations. Our model belongs to the class of ``on-off'' models discussed
in a seminal paper by Doering and Gadoua \cite{doe} and further analyzed by
Bogu\~n\'a \textit{et al.} \cite{boguna}. A distinctive characteristics of this model is
that part of the time the barrier is either switched off (\textit{i.e.} it becomes 
flat), or
the switching is performed between the barrier and a well, so that the particle
can essentially roll rather than climb during these times.

\begin{figure}[!h]
\includegraphics[angle=0, width=8.5cm, height=8.5cm]{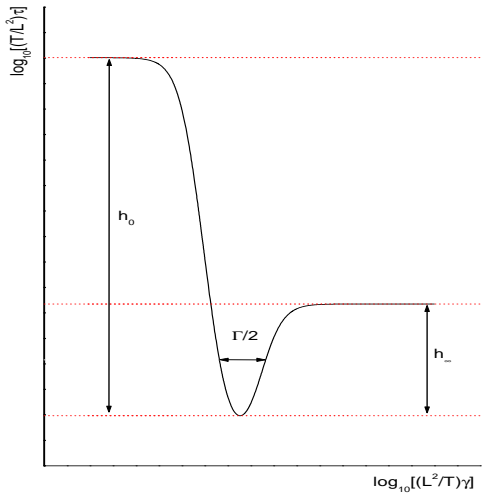}
\caption{\label{lin}The resonant $\tau(\gamma)$ line  and its asymptotes at low and high external
noise frequencies.}
\end{figure}

Our major goal was to examine variability of the mean first passage time in
parameter space, that is, to determine 
the position
of its minimum (\textit{cf.} Fig.~\ref{lin}) and to study the dependence of depth and width of this minimum on
the parameters describing the barrier shape and correlation time of the
environmental noise.

In Section II, a brief statement of the ``archetypal'' resonance activation
problem is presented.  Section III discusses solutions to the problem posed for
the 
\textit{arctan} potential, pointing out special symmetry of this function that 
determines equality of the mean first passage time for convex and concave slopes
approximating between the triangular, piecewise-linear and piecewise-constant 
 forms of the potential. Numerical
results are obtained and further analyzed to asses how sharp and persistent
the resonant region of the mean first passage time can be as a function of the
noise correlation time and steepness of the potential slope. We conclude with
some final remarks in Section IV.

%
%
\section{Generic model system}
We have considered an overdamped Brownian particle
\cite{kra}  moving in a potential field
between absorbing and reflecting boundaries in the presence of noise
that modulates the barrier  height. 
The evolution of a state variable $x(t)$ is described in terms of the 
Langevin equation
\begin{eqnarray}
\frac{dx}{dt} & =& -V'(x)+\sqrt{2T}\xi(t)+g(x)\eta(t) \nonumber \\
&= & -V_\pm^{'}(x)+\sqrt{2T}\xi(t).
\label{lang}
\end{eqnarray}
Here $\xi(t)$ is a Gaussian process with zero mean and correlation
 $<\xi(t)\xi(s)>=\delta(t-s)$ (\textit{i.e.} the Gaussian white noise arising
 from the heat bath of temperature $T$),
$\eta(t)$ stands for  a Markovian dichotomous (not necessarily symmetric)
 noise taking on one of
two possible values $a_\pm$ and prime means differentiation over $x$. The correlation function of the dichotomous process satisfies
$<\eta(t)\eta(t')>=e^{-2\gamma |t-t'|}{(a_{+}-a_{-})^2}/4$
where $\gamma$ stands for  the flipping frequency
of the barrier fluctuations  (\textit{i.e.} $\frac{1}{2\gamma}$ represents the characteristic correlation time of the dichotomous noise). 
Both noises are assumed to be statistically independent, \textit{i.e.} 
$<\xi(t)\eta(s)>=0$.
Equivalent to eq.~(\ref{lang}) is a set of the Fokker-Planck equations 
describing evolution of the probability density of finding the particle
 at time $t$  at a position $x$, subject to the force 
 $-V_\pm^{'}(x)=-V'(x)+a_\pm g(x)$

\begin{eqnarray}
\partial_t {P}(x,a_\pm,t)& =&  \partial_x  \left[V_\pm^{'}(x)+T\partial_x  \right]P(x,a_\pm,t) \nonumber \\
  & -& \gamma P(x,a_\pm,t)+\gamma P(x,a_\mp,t) 
\label{schmidr}
\end{eqnarray}
In the above equations time has dimension of $[length]^2/energy$ due to a friction
constant that has been ``absorbed'' in a time variable.
We are assuming a  reflecting
boundary at $x=0$ and an absorbing
boundary condition at $x=L$
\begin{equation} 
P(L,a_\pm,t)=0
\label{bon0}
\end{equation}
\begin{equation}
\left[V_\pm^{'}(x) +T\partial_x\right]P(x,a_\pm,t)|_{x=0}=0.
\label{bon}
\end{equation}
The initial condition
\begin{equation}
P(x,a_+,0)=P(x,a_-,0)=\frac{1}{2}\delta(x)
\end{equation}
expresses equal choice to start with any of the two configurations of the
barrier.
The quantity of interest is the mean first passage time 
\begin{eqnarray}
\tau & = &\int\limits_0^\infty
dt\int\limits_0^L\left[P(x,a_+,t)+P(x,a_-,t)\right]dx \nonumber \\
&=&\tau_+(0)+\tau_-(0)
\end{eqnarray}
with $\tau_+$ and $\tau_-$ being MFPT for $(+)$ and $(-)$ 
configurations, respectively. MFPTs $\tau_+$ and $\tau_-$
 fulfill the set of backward Kolmogorov equations \cite{goel,bork} 
\begin{eqnarray}
-\frac{1}{2} & = &  - \gamma\tau_\pm (x)+\gamma\tau_\mp(x) \nonumber \\
& - & \frac{dV_\pm (x)}{dx}\frac{d\tau_\pm (x)}{dx}+T\frac{d^2\tau_\pm (x)}{dx^2}
\label{mr_uklad}
\end{eqnarray}
with the boundary conditions (\textit{cf.} eq.~(\ref{bon0}) and~(\ref{bon}))
\begin{equation}
\tau_{\pm}^{'}(x)|_{x=0}=0
\qquad
\tau_{\pm}(x)|_{x=L}=0
\end{equation}
Although the solution of~(\ref{mr_uklad}) is usually unique \cite{molenaar},
 a closed, ``ready to use'' analytical formula for $\tau$ can be obtained only 
 for the simplest cases of
 the potentials (piecewise linear and piecewise constant). More complex cases, like even piecewise 
 parabolic potential $V_\pm$ result in an intricate form of the solution to
 eq.~(\ref{mr_uklad}).
 Other situations require either use of approximation schemes \cite{rei},
 perturbative approach \cite{iwa} or direct numerical evaluation methods \cite{gam}.
In order to  examine MFPT  for various potentials, a modified program \cite{musn}
applying general shooting methods has been used with part of the mathematical
software obtained  from the \textit{Netlib} library.

%
%

\section{Solution and Results}
For $\gamma=0$ and 
$V_+(x)=V_-(x)=V(x)$, eq.~(\ref{mr_uklad}) describes an overdamped Brownian particle moving in an external static potential 
(\textit{cf.} eq.~(\ref{lang}) with $\eta(t)\equiv 0$).
In this situation a formula for the MFPT reads
\begin{eqnarray}
\tau(0\to L)&=&\frac{1}{T}\int\limits_0^Ld\eta\exp\left(\frac{V(\eta)}{T }\right)\int\limits_0^\eta\exp\left(-\frac{V(\xi)}{T}\right)d\xi
\nonumber \\
&=&\frac{1}{T}\int\limits_0^L\int\limits_0^L\exp\left(\frac{V(\eta)-V(\xi)}{T}\right)\theta(\eta-\xi)d\eta d\xi.
\end{eqnarray}
and does not depend on a constant part of a potential.
Another non-trivial property in this case is equality of times
\begin{equation}
\tau(U,0\to L)=\tau(V,0\to L),
\label{property}
\end{equation}
true for any two potentials $V(x)$, $U(x)$ that fulfill
\begin{equation}
U(x)=\mathcal{H}-V(L-x)
\label{cos}
\end{equation}
and are continuous functions mapping $[0,\;L]\to[0,\;\mathcal{H}]$.
The above  property can be simply proven after noticing that the difference
\begin{equation}
\tau(V,0\to L)-\tau(U,0\to L)  = \int\limits_0^L\int\limits_0^Lf(\eta,\xi)d\eta d\xi 
\end{equation}
with
\begin{eqnarray}
f(\eta,\xi) & = & \left[\exp\left(\frac{V(\eta)-V(\xi)}{T}\right)-\exp\left(\frac{U(\eta)-U(\xi)}{T}\right)\right] \nonumber \\
& \times & \frac{\theta(\eta-\xi)}{T}
\end{eqnarray}
vanishes for any points $(p,q)$ and $(p',q')$ such that $(p',q')$ is the image of
$(p,q)$ in the reflection at the line $\eta=L-\xi$
\begin{equation}
f(p,q)=-f(p',q').
\label{property2}
\end{equation}

\begin{figure}[!h]
\includegraphics[angle=0, width=7.0cm, height=8.5cm]{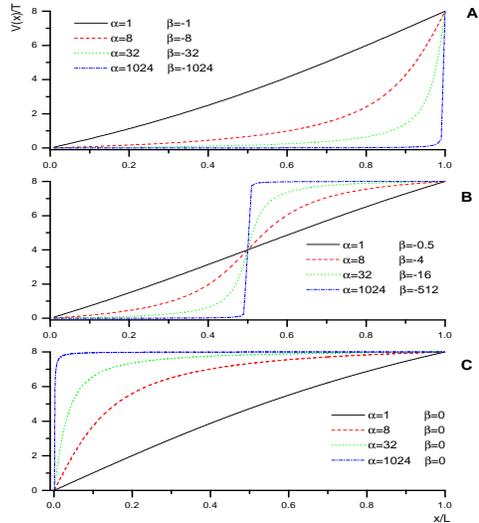}
\caption{\label{pot}Potentials of \textit{arctan} type for: \textbf{A} $\beta =-\alpha $, \textbf{B} $\beta =-\alpha/2$, \textbf{C} $\beta
=0$.}
\end{figure}

To capture the features of the mean first passage time as a function of the shape
of the barrier and characteristic correlation time of the barrier fluctuations,
we have studied the Brownian diffusion problem in the \textit{arctan} 
potential fluctuating randomly between two configurations $V_+(x)$ and $V_-(x)$
\begin{eqnarray}
V_\pm(x)& =& \frac{H_\pm \mbox{arctan}(\alpha x/L+\beta )}{ \mbox{arctan}(\alpha +\beta )-\mbox{arctan}(\beta )} \nonumber \\
 & - & \frac{H_\pm\mbox{arctan}(\beta )}{\mbox{arctan}(\alpha +\beta )- \mbox{arctan}(\beta )}.
\label{pott}
\end{eqnarray}
Since switching between both configurations is modelled by a Markovian two-state
process, the potential experienced by a particle remains in a given
configuration for an exponentially distributed time, before flipping to the
other state.
 The 
change in steepness and shift of the potential slope is controlled by 
parameters $\alpha $ and $\beta $. By using  this particular form of the potential,
we are able to modulate the steepness of the slope when holding the same, constant 
values of
$ V_\pm(x)$ at the bottom of the potential ($x=0,V_\pm(0)=0 $) and at the top 
($x=L,V_\pm(L)=H_\pm$) of the barrier
 (see Fig.~\ref{pot}). Moreover, as it is clearly seen from eq.~(\ref{pott}), for
 $\beta=0$  and $\beta=-\alpha$, the
 potential satisfies the property eq.~(\ref{cos}) with $\mathcal{H}=H_{\pm}$
\begin{equation}
V_\pm(x)|_{\beta=0}=H_\pm -V_\pm(L-x)|_{\beta=-\alpha}.
\end{equation}
Thus, under the same type of barrier flipping process, the MFPT results 
  are identical for the convex ($\beta=0$)
 and concave ($\beta=-\alpha$) potential slopes.

The standard numerical analysis \cite{molenaar,rec} of the MFPT has been performed 
for various forms of the potential eq.~(\ref{pott}) 
with the slope changing accordingly to  $\alpha =2^i$, $i=0,\dots,10$ and $\beta =-\alpha $
(Fig.~\ref{pot}~A),
$\beta =-\alpha /2$ 
(Fig.~\ref{pot}~B) and  $\beta =0$ (Fig.~\ref{pot}~C). The height of the potential barrier $H_\pm$ has been 
switching between either two symmetric
values,
$H_\pm=\pm8T$, or ``on'' and ``off'' barrier positions, 
\textit{i.e.} $H_+=8T,\;H_-=0$.
Numerical analysis 
 reestablishes discussed above equality of the MFPTs for potentials
with $\beta=-\alpha$ and $\beta=0$. It means that for this set of parameters, 
the property eq.~(\ref{property}) 
observed for static potential, is also recovered 
 for dichotomously fluctuating barriers
\begin{equation}
\tau(U,\gamma)=\tau(V,\gamma)
\label{pro}
\end{equation}
Therefore, graphical presentation of the results relates
arbitrarily to the convex potentials only ($\beta=0$), having in mind that 
the same relationship  $\tau(\gamma)$ is observed for the concave
potentials with $\beta=-\alpha$ (\textit{cf.} Fig.~\ref{pot}).

Generally with increasing  slope of the barrier, resonant frequencies are 
shifted to higher values suggesting that more energy has to be pumped into the
system in order to facilitate the escape of a Brownian particle over the fluctuating barrier.
The minimal (resonant) values of $\tau$  increase
with $\alpha $ for $V_\pm=\pm V$ potential and decrease for potentials switching
between the ``on-off'' ($V_+=V,\;V_-=0$) positions, thus documenting a  different
character of resonant activation in these two cases, as discussed by Bogu\~n\'a
 \textit{et al.} \cite{boguna}.

\begin{figure}[!h]
\includegraphics[angle=0, width=8.5cm, height=8.5cm]{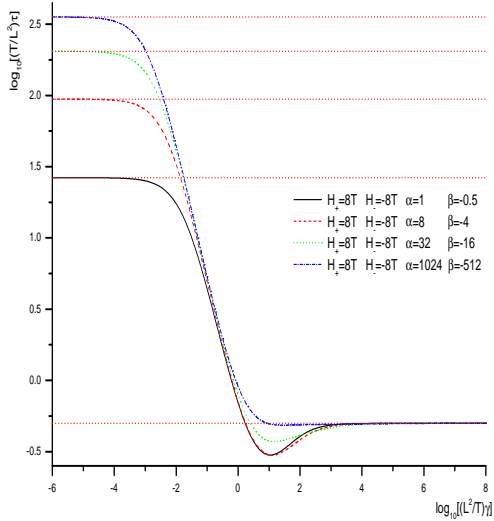}
\caption{\label{st05sk}The $\tau(\gamma)$ as a function of barrier fluctuation rate for \textit{arctan}
 potentials with $H_+=8T,\;H_-=-8T$, $\beta =-\alpha /2$.}
\end{figure}
\begin{figure}[!h]
\includegraphics[angle=0, width=8.5cm, height=8.5cm]{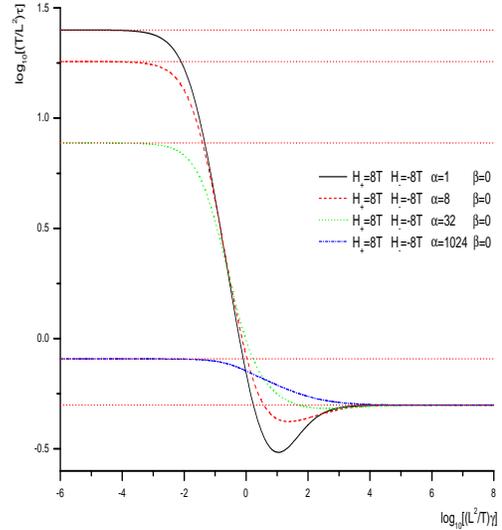}
\caption{\label{st0sk}The $\tau(\gamma)$  as  a function of barrier fluctuation rate for \textit{arctan}
 potential with $H_+=8T,\;H_-=-8T$, $\beta =0$ ($\beta=-\alpha$).}
\end{figure}

\begin{figure}[!h]
\includegraphics[angle=0, width=8.5cm, height=8.5cm]{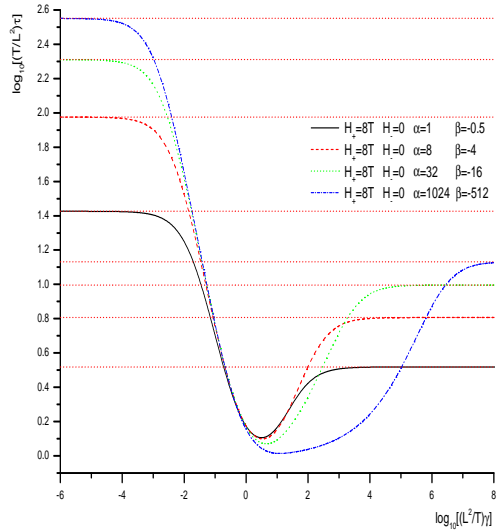}
\caption{\label{at05sk}The $\tau(\gamma)$  as a function of barrier fluctuation rate for 
\textit{arctan} potentials with $H_+=8T,\;H_-=0$, $\beta =-\alpha /2$.}
\end{figure}

\begin{figure}[!h]
\includegraphics[angle=0, width=8.5cm, height=8.5cm]{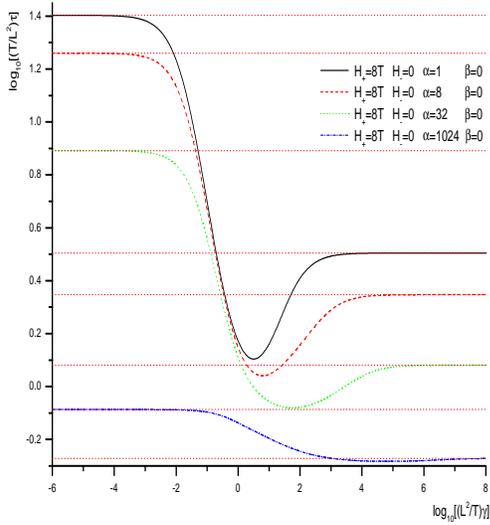}
\caption{\label{at0sk}The $\tau(\gamma)$  as a function of barrier fluctuation rate for \textit{arctan}
 potential with $H_+=8T,\;H_-=0$, $\beta =0$ ($\beta=-\alpha$).}
\end{figure}

\begin{figure}[!h]
\includegraphics[angle=0, width=8.5cm, height=8.5cm]{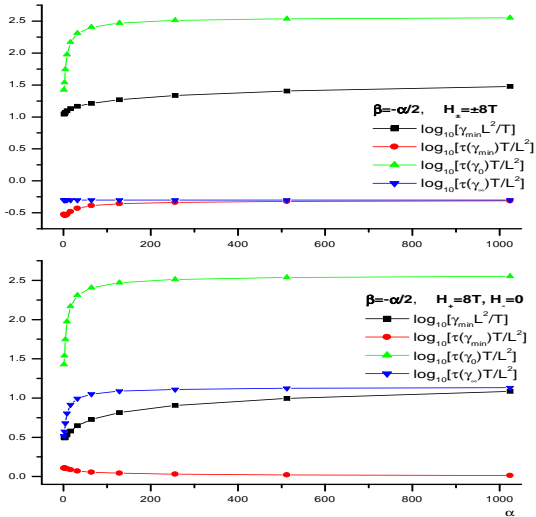}
\caption{\label{tabsk2}Location of minima, minimal and asymptotic values of the 
MFPT for \textit{arctan} type potentials with $\beta=-\alpha /2$. Lines have
been drawn to guide the eye.}
\end{figure}

\begin{figure}[!h]
\includegraphics[angle=0, width=8.5cm, height=8.5cm]{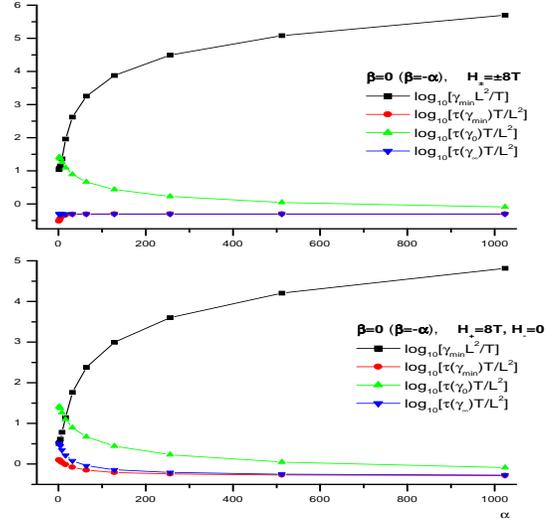}
\caption{\label{tab0}Location of minima, minimal and asymptotic values of the MFPT for \textit{arctan}
 type potentials with $\beta =0$ ($\beta=-\alpha$). Lines have
been drawn to guide the eye.}
\end{figure}

For $V_\pm=\pm V$ barriers asymptotic values of MFPT at $\gamma\rightarrow\infty$
 do not change with the steepness:
they remain the same (Figs.~\ref{st05sk},~\ref{st0sk}) and equal to
$\frac{L^2}{2T}$, as predicted
by analytical results for piecewise linear and piecewise constant potentials \cite{doe,zucher,boguna}. However, they alter for the
``on-off''
potentials  (\textit{cf.} Figs.~\ref{at05sk} and~\ref{at0sk}) displaying increase for $\beta =-\alpha /2$  and decrease for
 $\beta =0$  (same for concave potentials $\beta=-\alpha$).
The resonant line (Fig.~\ref{lin}) of the $\tau(\gamma)$ dependence 
displays different
features with zero and non-zero shift $\beta $. In particular,  for $\beta =0$
the relative depths of minima  measured from either the asymptotic $\tau$
 value at $\gamma=0$  ($h_0$), or from the
$\tau$ asymptote 
at $\gamma\rightarrow\infty$  ($h_{\infty}$) decrease with increasing $\alpha$ (\textit{cf.} Figs.~\ref{st0sk} and~\ref{at0sk}).
For non-zero $\beta =-\alpha /2$ and both potentials, $V_\pm=\pm V$ or  $V_+=V,\;V_-=0$, the
 relative depth $h_0$  increases with $\alpha $,  as documented in Fig.~\ref{tabsk2}
where the  
 distance between the lines $\log_{10}[\tau(\gamma_0)T/L^2]$ and
 $\log_{10}[\tau(\gamma_{min})T/L^2]$ is shown to  
increase with this parameter.
At the same conditions, the depth of the resonant line $h_{\infty}$ measured from the
asymptote at $\gamma\rightarrow\infty$ decreases for $V_\pm=\pm V$ potentials and
increases for the ``on-off'' potentials $V_+=V,\;V_-=0$.
These findings are summarized in Figs.~\ref{tabsk2} and~\ref{tab0} where
logarithmic values of $\gamma_{min}$, $\tau(\gamma_{min})$, 
$\tau(\gamma_{0})$, $\tau(\gamma_{\infty})$
 for the all \textit{arctan} type potentials are plotted \textit{versus} varying slope parameter $\alpha $.


Moreover, as can be easily seen in Figs.~\ref{at05sk} and~\ref{at0sk}, the resonant width $\Gamma/2$
 measured at $h_{\infty}/2$  (\textit{cf.} Fig.~\ref{lin}) increases significantly 
for the ``on-off'' potentials with increasing value of the slope parameter $\alpha $.
It is thus suggestive of an easier fine-tuning of the system to resonant
conditions in this case by switching frequency of the barrier.
The effect is also reproducible for $\beta =0$, although in this case the relative depth of 
MFPT minimum decreases, causing observable ``flattening'' of the resonant
line.


The functional dependence  $\tau(\gamma)$ has a typical character in all
 cases discussed in this work.
In agreement with theoretical considerations
\cite{doe,zucher,pechukas,rei,reimann,schmid,hanggi}, the asymptotic values 
$\tau(\gamma_0)$  and $\tau(\gamma_{\infty})$ for large and small correlation 
times of the
barrier noise  are recovered with a minimum located in between those two.
For all  slopes of barriers this characteristic behavior of the MFPT remains similar
although the weakening of the effect, \textit{i.e.} shift of the resonant
$\gamma_{min}$ to higher values and shallowing of the resonant line are 
observed with increasing steepness of the barrier.
The effect of attenuation is stronger for $V_\pm=\pm V$ potentials with both,
 $\beta =-\alpha /2$ or $\beta =0$ ($\beta=-\alpha$). The resonant line flattens
 and slowly disappears with increasing $\alpha$ (\textit{i.e} with increasing
 steepness of the barrier slope). The result is documented in Figs.~\ref{tabsk2} (upper
 panel) and~\ref{tab0}, where the noticeable convergence between the values 
 $\tau(\gamma_{min})$ and $\tau(\gamma_{\infty})$ is observed for $\alpha=1024$.
 Numerical estimates of MFPTs for the alternating potential ($V_\pm=\pm 8T$)  yield
 $\tau(\gamma_{min})\approx 0.3$ and $\tau(\gamma_{\infty})\approx 0.5$
  for $\alpha=8, \beta=-4$ leading to $h_{\infty}\approx 0.2$. This value drops 
  by an order of magnitude to
  $h_{\infty}\approx 0.01$ for $\alpha=1024, \beta=-512$, that corresponds to
   by two-order of magnitude higher steepness of the slope estimated
  as the derivative of the potential at point $x=L/2$. Thus, the result suggests
  a fairly robust character of the resonant activation that disappears, but
  only for the limiting step-function potentials. An apparent persistence of the
  phenomenon for $V_+=8T, V_-=0$ and $\beta=-\alpha/2$ (Figs.~\ref{at05sk} and~\ref{tabsk2}, lower panel) is due to a different
  scenario of the barrier passage. In this particular case, an increasing
  $\alpha$ produces a wall at $x=L/2$  which has to be passed by a Brownian
  particle before being eventually absorbed at $x=L$. The particle experiences then
 a non-zero deterministic force at this point, only. 

Potentials of the \textit{arctan} type (eq.~(\ref{pott}) and Fig.~\ref{pot}) produce 
barriers approximating between triangular and rectangular shapes mostly used for
analytical studies of the resonant activation effect
\cite{doe,bier,zucher,boguna}. Although the
resonant phenomenon has been detected and analytically proven as a generic property in the case of
dichotomously switching piecewise linear potentials \cite{doe,zucher}, it has
been shown \cite{zucher} not to 
occur in the case of piecewise constant potentials, when the Brownian particle escapes
over either the small or the large barrier at all times.
In consequence of the latter, barrier crossing events and
barrier fluctuations  are two independent stochastic processes
\cite{zucher,broeck,reihan}  with the mean exit time
 decreasing monotonically from $\tau\approx \tau_+$ for $\gamma< \tau_+^{-1}$ to
$\tau\approx \tau_-$ for $\gamma>\tau_-^{-1}$. The qualitative behavior of
$\tau$ is then identical with the behavior of the exit time in so called kinetic
models \cite{zucher,broeck,reimann,pechukas,gaveau} which explicitly include the escapes over both high and low barriers at
a rate $W_{\pm}=\exp(-H_{\pm}/T) $ and
transitions between the two configurations of the potential. As it has been
discussed by Bier and Astumian \cite{bier}, the kinetic
description assumes an instantaneous adiabatic adjustment 
and accordingly leads
to the set of equations for the populations $\pi_{\pm}$ of  particles that
feel the potential $V_{\pm}$ and have not yet escaped over the
barrier
\begin{equation}
\frac{d}{dt}\pi_{\pm}(t)=-W_{\pm}\pi_{\pm}-\gamma \pi_{\pm} + \gamma \pi_{\mp}
\end{equation}
 The decay of the 
survival probability $\pi(t)=\pi_{+}(t)+\pi_{-}(t)$ is then determined
by an average MFPT 
\begin{equation}
\tau \equiv \tilde{T}=-\int_0^{\infty}t\frac{d\pi(t)}{dt}dt=\frac{2\gamma+(W_{+}+W_{-})/2}
{W_{+}W_{-}+(W_{+}+W_{-})\gamma}
\end{equation}
which for slow barrier fluctuations ($\gamma=0$) approaches the average of 
the MFPT for the alternative 
configurations and in the fast fluctuation limit ($\gamma \rightarrow\infty$ but
respecting $\gamma<<T_s^{-1}$, where $T_s^{-1}$ stands for the frequency of the
escape attempts \cite{rei,reimann,reihan})
coincides with the reciprocal of the average rate over the fluctuating
barrier. 
For fluctuating rectangular potentials with a barrier placed at
$x=L/2$, the closed expressions for the mean
 escape time \cite{zucher} for $\gamma=0$ and $\gamma\rightarrow\infty$ read
\begin{equation}
\tau (\gamma_0) \frac{T}{L^2} =\frac{1}{2}\left[\frac{\tau_+
+\tau_-}{2}\right],
\end{equation}
\begin{equation}
\tau (\gamma_{\infty}) \frac{T}{L^2}
=\frac{1}{2}\left [1+\frac{\tau_+\tau_--1}{2+\tau_++\tau_-}\right ]
\label{infty}
\end{equation}
and are perfectly reproduced in numerical MFPT studies in the $V_{\pm}=\pm V$
potential with parameters $\alpha=1024, \beta=-\alpha/2$ approximating a
step-like function $\theta(x-L/2)$. However, in the case of the ``on-off''
potential $V_+=8T, V_-=0$ and the same set of parameters $\alpha$, $\beta$
numerical $\tau$ value for $\gamma\rightarrow\infty$ differs from the
eq.~(\ref{infty}) by four order of magnitude (\textit{cf.}
Figs.~\ref{tabsk2},~\ref{at05sk}) displaying a persistent resonant activation
vanishing slowly at much steeper barriers which eventually reach the limit of an
infinite tangent at $x=L/2$. 
Similarly to the former results \cite{zucher}, in the limiting case of a
piecewise constant potential, the mean exit time approaches $\tau\approx\tau_+$
for $\gamma<\tau_+^{-1}$ and $\tau\approx\tau_-$ for $\gamma>\tau_-^{-1}$
displaying an inverse proportionality to the rate of the barrier fluctuations
for intermediate rates.

%
%
\section{Summary}
In the foregoing sections we have considered the thermally activated
process that occurs in a system coupled to an external noise source. The external
stochastic process is responsible for fluctuations of the potential barrier which has
been modelled by an \textit{arctan} function of varying slope.
As expected based on previous theoretical studies \cite{doe,rei}, the phenomenon of
resonant activation occurs typically in the system under broad circumstances of 
varying shape of the potential barrier and qualities of barrier fluctuations.
In general, with  increasing barrier steepness the resonance phenomenon 
becomes less ``sharp''
with the flat region of flipping frequencies around the resonant one. 
The time scale of the resonant activation process is fairly insensitive to the potential
parameters except in the case of the ``on-off'' potentials, when broadening of the MFPT
resonant line occurs suggesting most efficient tuning of the system to the resonance
conditions.

\begin{acknowledgments}
The authors express their gratitiude to the Referee whose critical comments
have helped to improve the manuscript.
This work has been partially supported by the Research Grant (1999-2002)
from the Marian Smoluchowski Institute of Physics.
\end{acknowledgments}

%
%

%
%

\end{document}